\documentclass[aip,preprint,amsmath,amssymb]{revtex4-2}

\usepackage[version=4]{mhchem}
\usepackage[colorlinks=true,urlcolor=blue,citecolor=blue]{hyperref}
\usepackage{graphicx}\graphicspath{{asset/}}
\usepackage{dcolumn}\usepackage{bm}

\usepackage[utf8]{inputenc}
\usepackage{etoolbox}
\usepackage{natmove} \usepackage{siunitx}
\usepackage{multirow}
\usepackage{changes}
\usepackage{float} 

\makeatletter
\def\@email#1#2{\endgroup
 \patchcmd{\titleblock@produce}
  {\frontmatter@RRAPformat}
  {\frontmatter@RRAPformat{\produce@RRAP{*#1\href{mailto:#2}{#2}}}\frontmatter@RRAPformat}
  {}{}
}\makeatother
\begin{document}

\preprint{AIP/123-QED}

\title{Electrostatic transfer of sub-micron magnetic particles onto cantilevers using a focused ion beam system}
\author{Peter Sun}
\email{hs859@cornell.edu}
\altaffiliation{Present address: Brookhaven National Laboratory, Upton, New York 11973, USA}
\author{George R. Du Laney}\affiliation{
Department of Chemistry and Chemical Biology, Cornell University, Ithaca, NY 14853, USA
}

\author{Tim M. Fuchs}
\affiliation{Leiden Institute of Physics, Leiden University, P.O. Box 9504, 2300 RA Leiden, The Netherlands}

\author{Tjerk H. Oosterkamp}
\affiliation{Leiden Institute of Physics, Leiden University, P.O. Box 9504, 2300 RA Leiden, The Netherlands}

\author{Malcolm G. Thomas}
\affiliation{Cornell Center for Materials Research, Cornell University, Ithaca, NY 14853, USA}

\author{John A. Marohn}\affiliation{
Department of Chemistry and Chemical Biology, Cornell University, Ithaca, NY 14853, USA
}

\date{\today}

\begin{abstract}
In this paper, we present a focused-ion-beam-assisted method for preparing magnet tips for magnetic resonance force microscopy measurements.
The method electrostatically transfers prefabricated magnetic nanoparticles to microcantilevers, achieving precise control over the magnet overhang past the cantilever leading edge while minimizing the fabrication damage to the leading edge of the tip magnet.
We demonstrate successful fabrication of magnets ranging in size from \SI{460}{\nano\metre} to \SI{2.8}{\micro\metre}. These magnets were affixed to two types of cantilevers: silicon cantilevers with a spring constant of \SI{800}{\micro\newton\per\metre}, and single-crystal silicon cantilevers with a spring constant of \SI{30}{\micro\newton\per\metre}.
We show that the electrostatic transfer method enables a wide variety of tip shapes, sizes, and materials that were previously not possible with conventional fabrication methods. The transfer procedure allows us to prefabricate the desired particle geometry with minimal ion-beam damage, as confirmed by Monte Carlo simulations. We show that the technique is versatile and can be used to fabricate custom-tipped cantilevers for a broader range of scanning probe techniques.
 \end{abstract}

\maketitle

\section{\label{sec:Introduction}Introduction}

The ability to reliably attach particles of known shape and material to cantilevers is of great interest in scanning probe microscopy (SPM).\cite{Mak2006apr,Gan2007aug}  Techniques have been developed for attaching microparticles\cite{Mak2006apr,Gan2007aug,Ducker1991sep}, nanoparticles\cite{Ong2007jun,Vakarelski2006mar,Stockle2000feb,Pettinger2004mar}, and single atoms and molecules\cite{Liu2012oct,Hwang2022sep} to probe tips for applications in atomic force microscopy\cite{Mak2006apr,Gan2007aug,Ducker1991sep,
Ong2007jun,Vakarelski2006mar,Liu2012oct}, tip-enhanced Raman spectroscopy \cite{Stockle2000feb,Pettinger2004mar,Wu2008}, scanning-tunneling microscopy \cite{Hwang2022sep}, and in magnetic resonance force microscopy (MRFM)\cite{Stipe2001mar,Jenkins2004may,Hickman2010dec}. This paper presents a novel approach to fabricating magnet-attached cantilevers for MRFM experiments. The approach allows fine control over magnet dimensions, material, and orientation, while minimizing particle damage.

Magnetic resonance force microscopy (MRFM) is a scanning probe technique that has imaged single electron spins in quartz \cite{Rugar2004jul}; detected magnetic resonance from small numbers of nitroxide spin labels \cite{Moore2009dec};  and has been proposed as a technique for determining, using spin labels, the three-dimensional location of proteins in biological complexes with angstrom precision \cite{Nguyen2018feb}.
In an MRFM experiment, a spin-containing sample is brought in proximity to a magnetic field gradient, usually generated by a nano-magnet\cite{Rugar2004jul,DeWit2019feb,Boucher2023jan,Sidles1995jan,Marohn1999oct}.
The work presented in this paper is motivated by an MRFM setup in which the nanomagnet is positioned at the tip of a microcantilever.
The magnet-on-cantilever MRFM experiment has been used to detect both electron spin resonance \cite{Wago1998may,Rugar2004jul,Moore2009dec, Vinante2011dec} and nuclear magnetic resonance \cite{Garner2004jun,Alexson2012jul,Wagenaar2016jul}.
Proposals and preliminary experiments have shown the feasibility of applying the MRFM technique to image nitroxide-labeled proteins \cite{Moore2009dec,Nguyen2018feb,Boucher2023jan}.

These studies, however, have found it challenging to achieve a high enough signal-to-noise ratio to detect single-electron spins in nitroxide radicals.
There are two conflicting requirements.
To increase the signal, we need a strong magnetic field gradient at the target spin; this gradient is maximized by operating at a small tip-sample separation.
At the same time, we must also reduce sample-induced force noise and frequency noise; we achieve this reduction by operating at large tip-sample separations, having the magnet overhang the leading edge of the cantilever, to decrease electrostatic interactions between trapped charges in the silicon cantilever and fluctuating electric fields and electric field gradients emanating from the sample \cite{Longenecker2012nov, Isaac2016}.
In practice, the signal-to-noise ratio is optimized by operating at an intermediate tip-sample separation with a magnetic tip whose radius is adjusted to maximize the field derivative at the given tip-sample separation \cite{Sidles1995jan,Kuehn2008feb}.
In small-tip experiments on nitroxide spins, Boucher and coworkers observed an electron spin signal that was 17 times smaller than calculated \cite{Boucher2023jan}.
Boucher hypothesized that this signal deficit was due to a combination of magnetization fluctuations and leading-edge magnet damage.
Thermally driven magnetization fluctuations in the nano-magnet could shorten the spin-lattice relaxation of sample electron spins \cite{Stipe2001mar,TseNgaNg2006mar}.
Thermal current fluctuations in the metallic magnet could create fluctuating magnetic fields that would also shorten spin-lattice relaxation \cite{Premakumar2018jan,Ariyaratne2018jun, Stipe2001dec, Kenny2021jun}.
Magnet damage during fabrication can reduce the magnet volume and therefore the tip's magnetic field gradient \cite{Overweg2015aug}.

To test these hypotheses, we need a magnet-tipped cantilever fabrication technique that produces magnets of different shapes, sizes, and materials (\emph{i.e.}, metallic, oxide, and alloy magnets) and affixes them to a cantilever.
Existing fabrication techniques do not meet these requirements.
Magnet-on-cantilever MRFM experiments employ extremely sensitive custom single-crystal silicon cantilevers made from silicon-on-insulator wafers \cite{Stowe1997jul, Jenkins2004may}.
In early work, a micron-scale magnet was glued to the cantilever using epoxy under an optical microscope \citep{Budakian2004jan, Moore2009dec} or under scanning electron microscopy (SEM) with nanopositioners \cite{Heeres2010feb}.
Larger magnets were milled to the desired shape and size using a focused-ion-beam (FIB) \cite{Budakian2004jan, Stipe2001dec}.
With this approach, the achievable magnet shape is limited because the cantilever is in the milling line of sight, and there is concern that FIB ions may damage the magnet's leading edge \cite{Stipe2001dec, Kato2004oct}.
Initial attempts to batch-fabricate magnets on cantilevers using electron-beam patterning, deposition, and lift-off were successful but had very low yield and produced relatively poor magnetic field gradients \cite{Hickman2010dec}.
Longenecker and coworkers fabricated the magnet in a separate chip using electron-beam patterning, deposition, and lift-off; the magnet chip was separated from the substrate using FIB milling and attached to a cantilever using FIB-induced deposition (FIBID) \cite{Longenecker2011may}; these tips produced record-high magnetic field gradients \cite{Longenecker2012nov}.
Others have grown magnetic nanoparticles on cantilevers using electron-beam-induced deposition \cite{Lavenant2014jan, Sangiao2017oct}.
Frustratingly, these fabrication approaches limit the magnet overhang, shape, size, material, and achievable tip magnetization.

In this paper, we present a new approach to serially fabricating magnets on cantilevers.
Here, the magnet is either prefabricated or milled by FIB into the desired shape, without exposing its leading edge to the ion beam.
The magnet is then transferred onto the cantilever using FIB and electrostatic forces, and attached to the cantilever using a platinum adhesion layer deposited by electron-beam-induced deposition (EBID).
This approach allows precise control of the magnet overhang and enables the creation of tip magnets from a broad range of shapes, sizes, and materials while minimizing adhesion material and beam damage. The procedure also allows optimized material pre-milling to reduce the milling damage.
We believe the technique can be applied to a broad range of mechanically detected SPM methods that require customized tips for detection.
 
\section{\label{sec:Methods}Methods}

\begin{figure}[!htbp]
    \centering
    \includegraphics[width=0.8\textwidth]{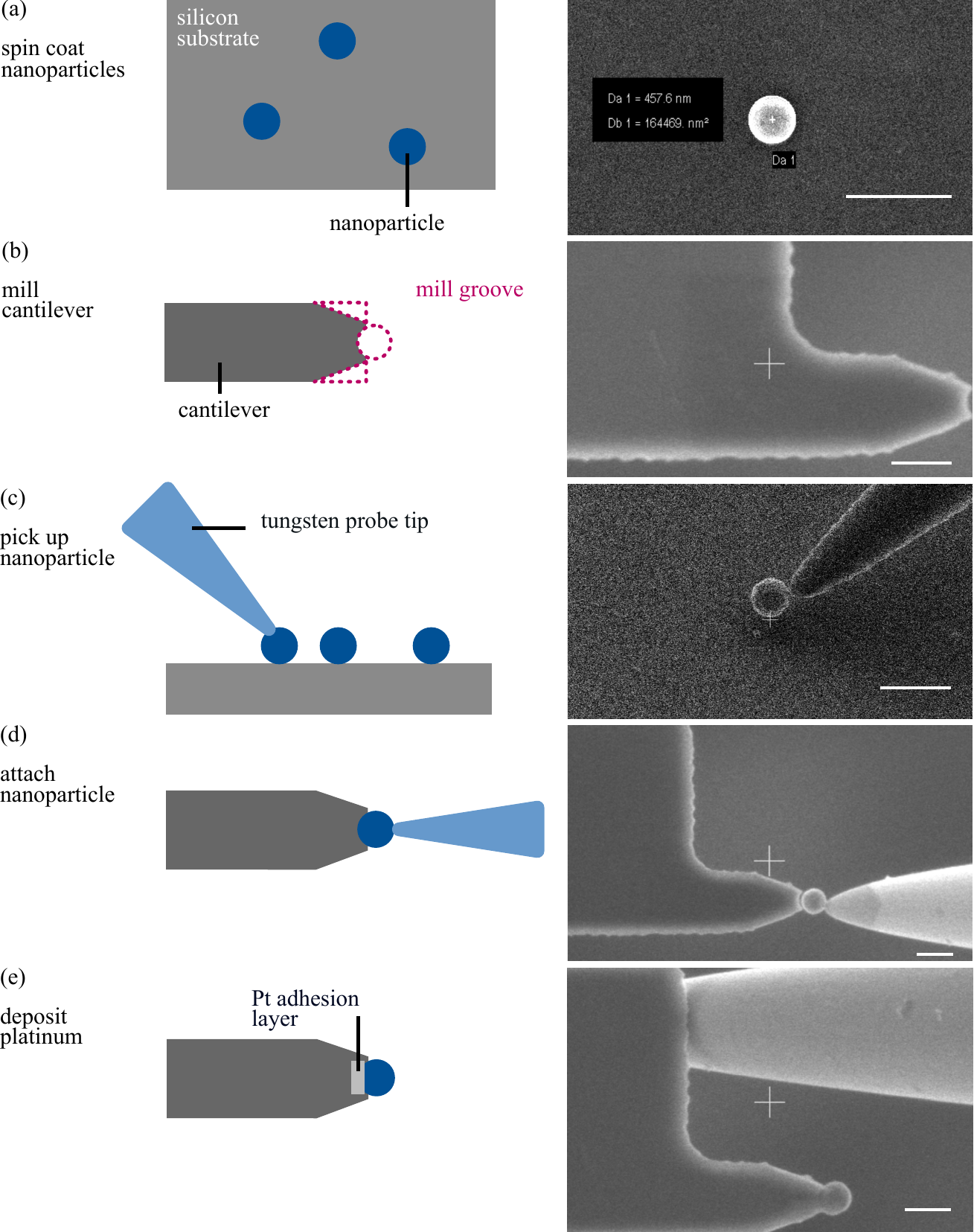}
    \caption{Electrostatic transfer of a \SI{460}{\nano\metre} Ni nanosphere onto a Type A MRFM cantilever. Left column: schematic. Right column: SEM images. (a) Spin-coated Ni particles on a silicon substrate. (b) Milled a cantilever notch at the leading edge. (c) Picked up the particle with the tungsten probe tip. (d) Attached the particle electrostatically to the cantilever. (e) Deposited Pt for adhesion, with a probe tip stabilizing the cantilever. Scale bars: \SI{1}{\micro\metre}.}
    \label{fig:fib-sphere}
\end{figure}

\begin{figure}[!htbp]
    \centering
    \includegraphics[width=0.8\textwidth]{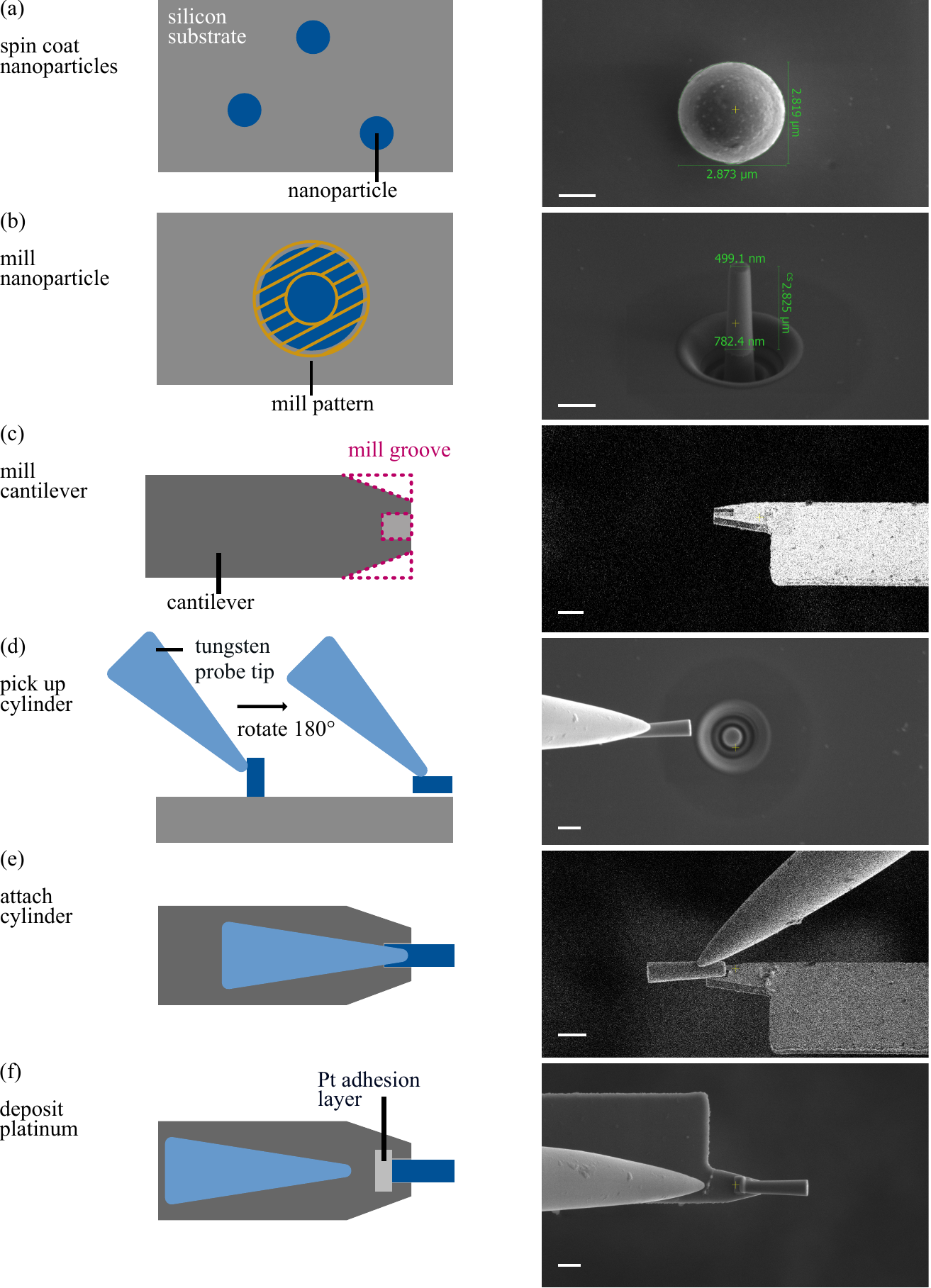}
    \caption{Electrostatic transfer of an FIB-milled NdFeB cylinder onto a Type A MRFM cantilever. Left column: schematic. Right column: SEM images. (a) Spin-coated NdFeB particles on a silicon substrate. (b) Milled a spherical NdFeB particle (\SI{2.8}{\micro\metre} diameter) into a sub-\SI{1}{\micro\metre}-diameter cylinder. (c) Milled a cantilever groove at the leading edge. (d) Picked up the particle with the tungsten probe tip. (e) Slotted the cylinder into the groove with the unexposed bottom facing away from the cantilever. (f) Deposited Pt by EBID to secure the magnet. The SEM image (b) was taken from a different transfer run from the rest of the SEM panels. Scale bars: \SI{1}{\micro\metre}. }
    \label{fig:fib-cylinder}
\end{figure}

The method consisted of magnet nanoparticle and cantilever preparation, followed by the electrostatic transfer and attachment.
We demonstrated this method using a \SI{460}{\nano\metre} diameter prefabricated Ni spherical magnet (Fig.~\ref{fig:fib-sphere}) and a \SI{2.8}{\micro\metre} long ion-milled NdFeB cylindrical magnet (Fig.~\ref{fig:fib-cylinder}). The procedures were performed using a Thermo Scientific Helios G4 UX focused ion beam (FIB) system. All SEM images of the transfer procedure were acquired using the FIB electron beam at \SI{5}{\kilo\electronvolt} and \SI{0.2}{\nano\ampere}.
The elemental content of the magnet nanoparticles was measured, before and after the transfer procedure, by energy-dispersive X-ray spectroscopy (EDS) on a Zeiss Gemini 500 scanning electron microscope (SEM) at an accelerating voltage of \SI{20}{\kilo\electronvolt}.

\subsection{Magnet Nanoparticle Preparation}

Nickel nanopowder (SkySpring Nanomaterials, Ni 99.5\%, 9210NG) and NdFeB powder (Magnequench, Nd-Pr-Fe-Co-Ti-Zr-B isotropic powder, MQP-S-11-9-20001) were suspended in isopropyl alcohol, sonicated, and filtered with appropriate Nylon syringe filters based on desired particle sizes (Tisch Scientific, \SI{1}{\micro\metre}, SPEC17986 and \SI{3}{\micro\metre}, SPEC18119). The suspension was spin-coated onto a silicon substrate at \SI{1500}{rpm} for \SI{30}{\second} to isolate the magnetic particles. Target isolated nanoparticles were located and measured using the scanning electron microscope (SEM) in the FIB, as shown in Fig.~\ref{fig:fib-sphere}(a) and Fig.~\ref{fig:fib-cylinder}(a). To obtain a cylindrical magnet, the NdFeB nanoparticle was milled using an annular pattern (Fig.~\ref{fig:fib-cylinder}(b)) with two-step milling first at \SI{30}{\kilo\electronvolt} and \SI{26}{\pico\ampere}, then at \SI{5}{\kilo\electronvolt} and \SI{7}{\pico\ampere}.

Two types of cantilevers were used.
Type A cantilevers (Figs.~\ref{fig:fib-sphere}, \ref{fig:fib-cylinder}, and \ref{fig:cylinder-redep}) were prepared from a commercial silicon-on-insulator wafer as described in Hickman et al.\ \cite{Hickman2010dec} and Jenkins et al.\ \cite{Jenkins2004may}.
These silicon cantilevers were \SI{400}{\micro\metre} long, \SI{5}{\micro\metre} wide, \SI{340}{\nano\metre} thick, and had a spring constant of \SI{800}{\micro\newton\per\metre}.
A Type B cantilever (Fig.~\ref{fig:ndfeb-transfer}) was prepared similarly, by Ben Chui in Dan Rugar’s group at IBM Almaden Research Center, and was obtained from the Oosterkamp Group at Leiden University.
The silicon cantilever was \SI{200}{\micro\metre} long, \SI{4}{\micro\metre} wide, \SI{100}{\nano\metre} thick, and had a spring constant of \SI{30}{\micro\newton\per\metre}.
For a spherical magnet attachment, a semicircular notch was milled to house the particle (Fig.~\ref{fig:fib-sphere}(b)). For a cylindrical magnet attachment, a rectangular groove was milled to house and align the particle (Fig.~\ref{fig:fib-cylinder}(c)).

\subsection{Electrostatic Transfer and Attachment}

A tungsten probe tip (Electron Microscopy Sciences, 75960-02) was used to pick up the selected magnet nanoparticle electrostatically. For spherical particles, the particle was attached to the bottom of the probe (Fig.~\ref{fig:fib-sphere}(c)). For cylindrical nanoparticles, the particle was attached to the side of the probe (Fig.~\ref{fig:fib-cylinder}(d)), and the probe was rotated 180 degrees to position the particle beneath the probe. The probe with the particle was then moved to the cantilever and slotted into the predefined notch. Electrostatic forces secured the magnet particles to the cantilever (Fig.~\ref{fig:fib-sphere}(d) and Fig.~\ref{fig:fib-cylinder}(e)).

The probe was then used to stabilize the cantilever during platinum deposition. The probe was pushed into the cantilever from the side (Fig.~\ref{fig:fib-sphere}(e)) or pushed down from the top (Fig.~\ref{fig:fib-cylinder}(f)). A final platinum adhesion layer was deposited using electron-beam-induced deposition (EBID) at \SI{5}{\kilo\electronvolt} and \SI{0.20}{\nano\ampere} to secure the magnet to the cantilever.
 
\section{\label{sec:Results}Results}

\begin{figure}[!htbp]
    \centering
    \includegraphics[width=\textwidth]{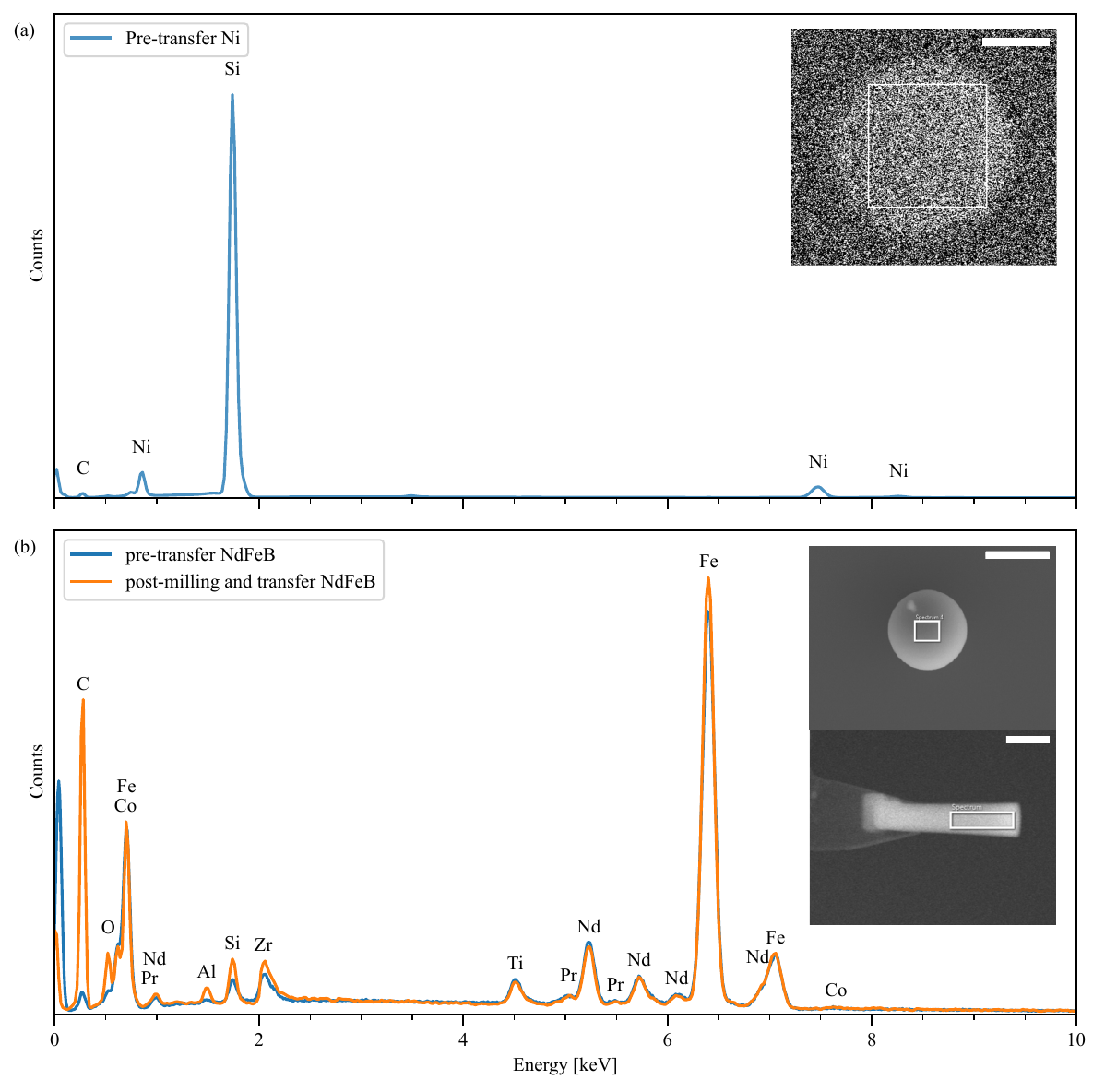}
    \caption{Energy dispersive X-ray spectroscopy (EDS) measurements of the magnetic nanoparticles before and after the transfer procedure. (a) Ni particle after spin coating, before transfer. (b) NdFeB magnet before and after milling and transfer. The inserted figures show the region of the EDS spectra. The Ni particle EDS post-transfer was not obtained because we do not expect significant changes in element contents. Scale bars for inserted images: (a) \SI{200}{\nano\metre}; (b) top: \SI{10}{\micro\metre}, bottom: \SI{1}{\micro\metre}. Both spectra were acquired at an accelerating voltage of \SI{20}{\kilo\electronvolt}, and only the \SIrange{0}{10}{\kilo\electronvolt} energy range is shown.}
    \label{fig:EDS}
\end{figure}

We successfully performed the electrostatic transfer on a spherical \SI{460}{\nano\metre} diameter nickel and a cylindrical \SI{2.8}{\micro\metre} long NdFeB alloy on \SI{800}{\micro\newton\per\metre} silicon cantilevers, as shown in the SEM images in Fig.~\ref{fig:fib-sphere} and Fig.~\ref{fig:fib-cylinder}. To confirm that we had the correct particle, an EDS spectrum of each nanoparticle after spin coating on a silicon substrate was measured; see Fig.~\ref{fig:EDS}. To investigate the effect of the milling and transfer procedure on the cylindrical magnet, we also measured the EDS spectra of the NdFeB magnet after the transfer procedure; see Fig.~\ref{fig:EDS}(b). The EDS spectra confirmed the elemental composition of the NdFeB cylinder before and after ion-beam milling and transfer; no significant compositional change was observed. The increased carbon, oxygen, and aluminum signals observed after transfer for the NdFeB particle are attributed to a reduced particle detection depth and to a change in the substrate from silicon to carbon tape on an aluminum stub.

\begin{figure}[!htbp]
    \centering
    \includegraphics[width=\linewidth]{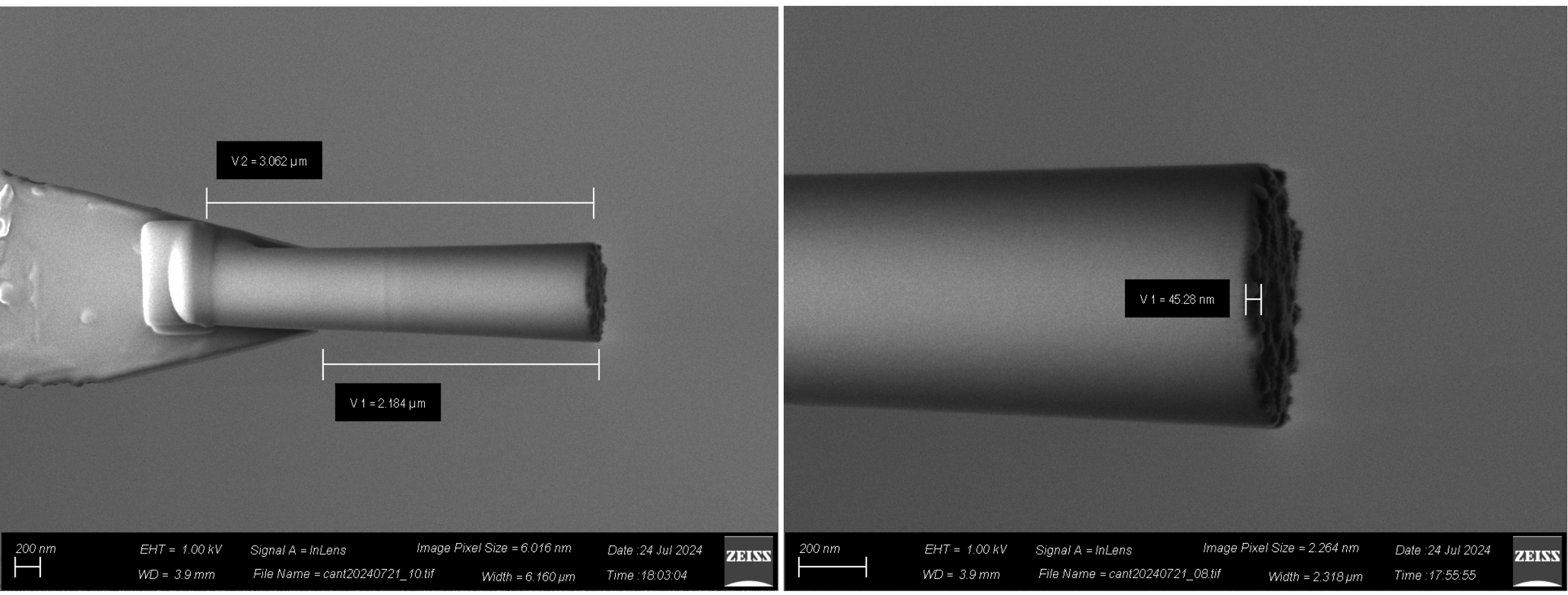}
    \caption{SEM images of cylindrical NdFeB magnet attached to Type A cantilever using the cylinder transfer protocol. The \SI{45}{\nano\metre} layer at the leading edge of the magnet is hypothesized to result from redeposition of NdFeB and silicon during milling. Scale bars: \SI{200}{\nano\metre}.}
    \label{fig:cylinder-redep}
\end{figure}

While attempting to minimize the damage of the leading edge, a closer inspection of the transferred cylindrical magnet revealed an additional \SI{45}{\nano\metre} layer at the leading edge of the magnet cylinder, hypothesized to be a mix of NdFeB and silicon material (Fig.~\ref{fig:cylinder-redep}). This additional layer was probably due to redeposition during milling, which filled the gap between the magnet's curved surface and the substrate.

\begin{figure}[!htbp]
    \centering
    \includegraphics[width=0.8\textwidth]{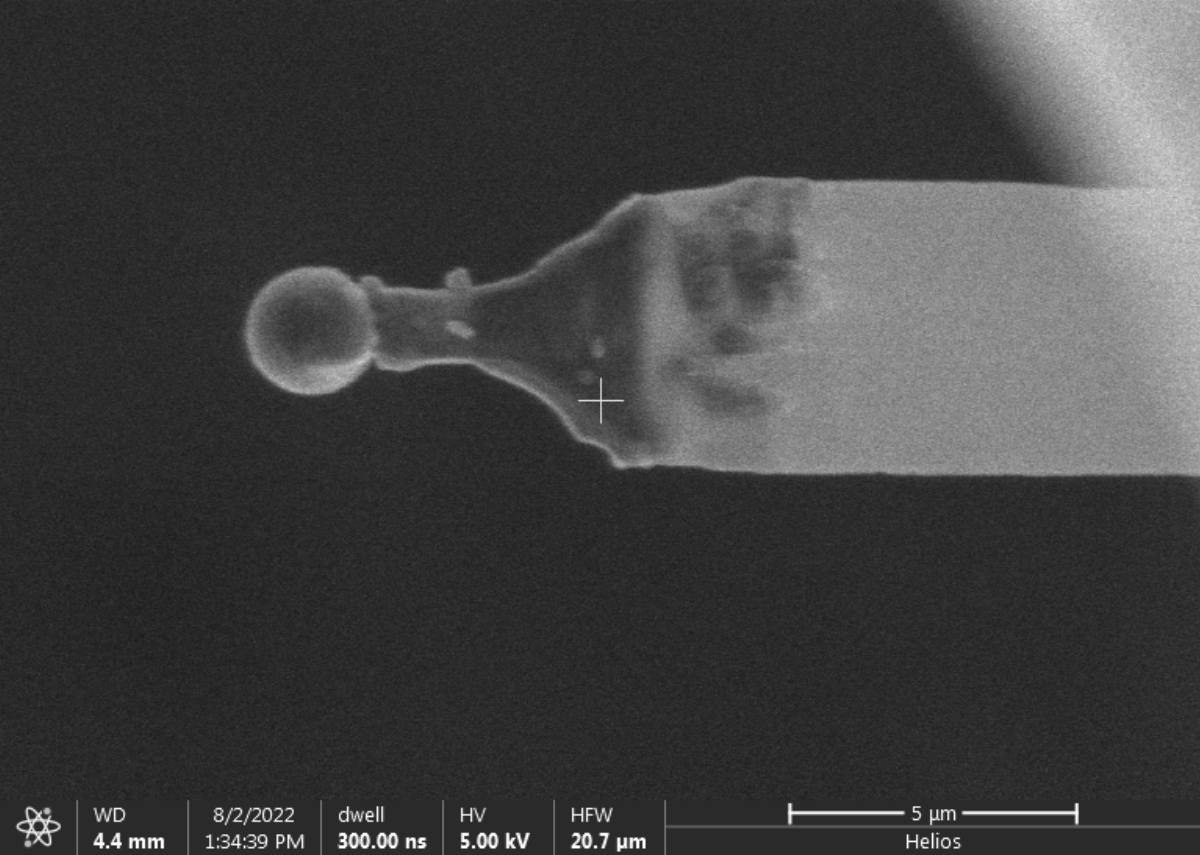}
    \caption{SEM image of spherical NdFeB magnet placed onto a Type B MRFM cantilever using the electrostatic transfer protocol. The NdFeB magnet diameter was \SI{2.2}{\micro\metre}. Scale bar: \SI{5}{\micro\metre}.}
    \label{fig:ndfeb-transfer}
\end{figure}

The Type A cantilevers in Figs.~\ref{fig:fib-sphere}, \ref{fig:fib-cylinder}, and \ref{fig:cylinder-redep} had a spring constant of only \SI{800}{\micro\newton\per\metre}, orders of magnitude lower than a commercial atomic force microscope cantilever.
We found that the transfer protocol could be used with an even more fragile cantilever.
In Fig.~\ref{fig:ndfeb-transfer} we show a \SI{2.2}{\micro\metre} diameter NdFeB spherical magnet affixed to a Type B cantilever; this cantilever had a spring constant of only \SI{30}{\micro\newton\per\metre}.

\section{\label{sec:Discussion}Discussion}

As discussed in the Introduction, achieving a high signal-to-noise ratio in MRFM requires meeting several key criteria. First, the magnet must overhang the cantilever leading edge to minimize surface-induced force noise from electrostatic interactions between trapped charges in the silicon and fluctuating fields from the sample.\cite{Hickman2010dec,Longenecker2012nov, Isaac2016} Second, the fabrication process must minimize damage to the magnet, particularly the leading edge closest to the sample spins.\cite{Overweg2015aug} Third, material, shape, and size flexibility are needed to test whether thermal magnetic fluctuations in ferromagnets\cite{Stipe2001mar,TseNgaNg2006mar} or thermal current fluctuations in metallic magnets \cite{Premakumar2018jan,Kenny2021jun} are causing the signal reduction observed in small-tip ESR experiments \cite{Boucher2023jan}. The FIB-enabled electrostatic transfer protocol introduced here addresses all these requirements for fabricating MRFM custom particle-tipped microcantilevers.

Rajupet et al.\ showed that, for particles interacting with rough surfaces, electrostatic charge dominates and changes slowly with particle distance. \cite{Rajupet2021apr} Similar behavior was observed during our transfer procedure, where magnet particles were attached to the probe tip. 
This behavior facilitated precise pick-up and attachment.
The custom notch or groove milled on the cantilever (Fig.~\ref{fig:fib-sphere}(b) and Fig.~\ref{fig:fib-cylinder}(c)) provided a slot to house the nanoparticle, ensuring the desired overhang distance (\SI{400}{\nano\metre} in Fig.~\ref{fig:fib-sphere}(b) and \SI{2}{\micro\metre} in Fig.~\ref{fig:fib-cylinder}(c)). The ion-beam-milled notch and groove became charged, facilitating electrostatic particle attachment. The milled groove for cylindrical particles ensured proper particle alignment during the final attachment step. The milling process also enabled control over the cantilever's cross-sectional area, which could decrease noncontact friction and frequency noise experienced by the cantilever.

The electrostatic transfer is geometrically agnostic. In particular, there is no restriction on the pickup point, and the probe tip can rotate to achieve the desired attachment angle. These degrees of freedom allow us to mill the nanoparticle into a wider range of geometric orientations than the glue-and-mill approach.\cite{Stipe2001dec, Overweg2015aug} Precise control over magnet shape enables optimization of magnet anisotropy and the magnetic field gradient profile to achieve higher sensitivity in MRFM experiments.\cite{Overweg2015aug, Campanella2011dec} Conventional direct fabrication approaches rely on deposition techniques such as e-beam evaporation or sputtering, which impose strict material constraints and require re-optimization for each new material. In contrast, transfer-based methods can directly utilize commercially available materials, removing these constraints and offering greater versatility with minimal optimization.
Unlike conventional approaches such as glue-and-mill, e-beam evaporation, and sputtering, which each impose constraints on material composition, particle geometry, or accessible size range, the electrostatic transfer protocol imposes no such constraints.

However, FIB milling can cause ion implantation, vacancy formation, and material redeposition, leading to magnetically inactive layers \cite{Faulkner2007dec, Fassbender2008feb, Overweg2015aug}. The redeposition can create tapered sidewalls.\cite{Lindsey2012jul, Desbiolles2019apr, Lid2021} This effect was evident in the cylindrical magnets produced here, where the final milled cylinder exhibited a \SI{2}{\degree} taper, and we observed redeposited milling material at the leading edge (Fig.~\ref{fig:cylinder-redep}). Ion-beam damage can cause magnetic volume reductions, which are undesired because they reduce the MRFM signal. In Ref.~\citenum{Overweg2015aug}, Overweg and coworkers fabricated a magnetic tip by milling at \SI{30}{\kilo\electronvolt} a FePt film into a magnet that was \SI{8.1}{\micro\metre} in length, \SI{1}{\micro\metre} in width, and \SI{300}{\nano\metre} in height. An estimated 30\% loss in magnetic moment for FePt magnets was observed. 

\begin{figure}[!htbp]
    \centering
    \includegraphics[width=1.0\textwidth]{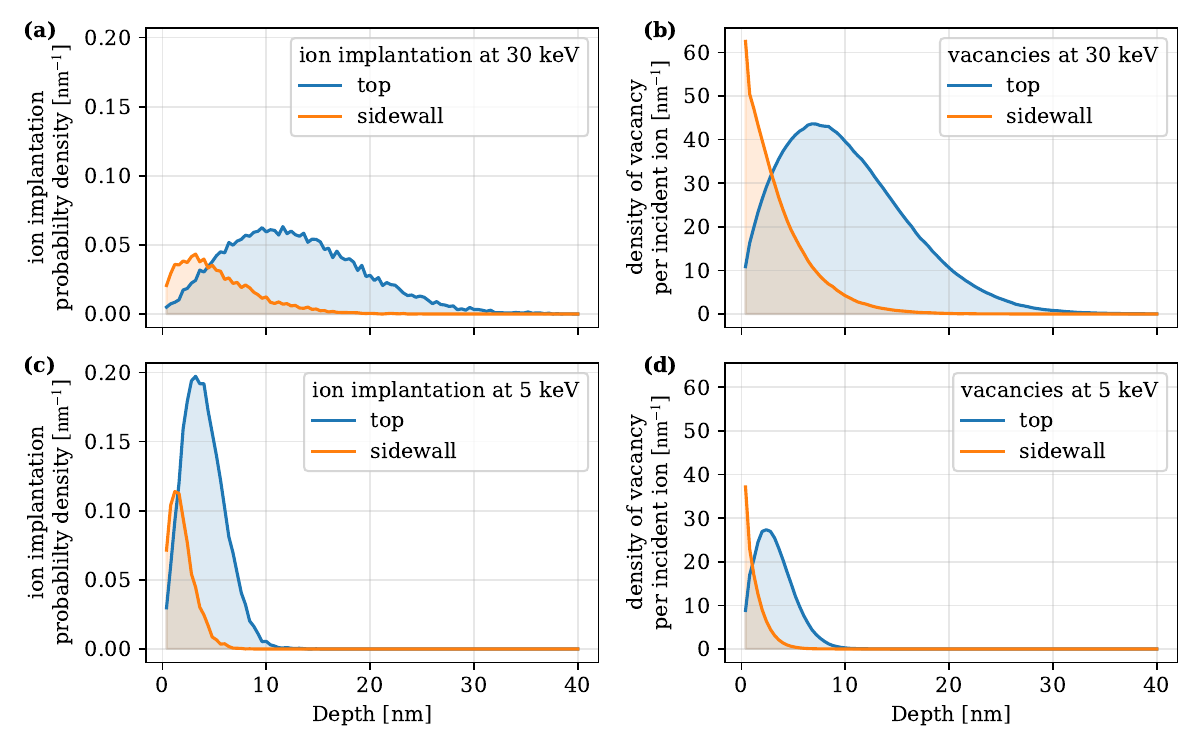}
    \caption{Stopping and Range of Ions in Matter (SRIM) Monte Carlo simulations of Ga$^+$ ion implantation into NdFeB. 
    Ion implantation probability density (a,c) and density of vacancies per incident ion (b,d) versus depth for \SI{30}{\kilo\electronvolt} (a,b) and \SI{5}{\kilo\electronvolt} (c,d) Ga$^+$ ions incident at \SI{0}{\degree} (top surface) and \SI{88}{\degree} (sidewall) from the surface normal.
    All profiles are normalized per incident ion per \si{\nano\metre} of depth, and a total of 20000 incident ions were simulated.
    In (a,c), the integrated area is $\leq 1$ because some ions are back-scattered.
    In (b,d), the integrated area is $\gg 1$ because 100s of vacancies are generated per incident Ga$^+$ ion. 
    Simulations were performed using SRIM-2013 \cite{Ziegler2010jun}.}
    \label{fig:fib-srim}
\end{figure}

The geometric freedom of the transfer procedure enables us to select the optimal milling angle, thereby further reducing ion-beam damage compared to conventional FIB milling approaches.
To mitigate ion-beam damage and the sidewall redeposition effect, we employed a two-pass milling approach: first milling at 30~keV to remove the bulk material, followed by a 5~keV milling pass to refine the shape, as suggested in Ref.~\citenum{Lid2021}. In our method, the magnet's body was exposed to the ion beam only at grazing incidence, and the leading edge was not exposed to the ion beam at all. Moreover, due to our two-pass approach and lower current, the tip receives a much smaller ion dose and damage depth than in Ref.~\citenum{Overweg2015aug}. In Fig.~\ref{fig:fib-srim}, the Stopping and Range of Ions in Matter (SRIM) simulation predicts that ion implantation and vacancy formation are significantly reduced at grazing angle (\SI{88}{\degree}) compared to normal incidence (\SI{0}{\degree}). At \SI{30}{\kilo\electronvolt}, the damage depth is estimated to be \SI{5.6 \pm 3.9}{\nano\metre} for the sidewall and \SI{12.8 \pm 6.4}{\nano\metre} for the top surface. The damage is significantly reduced at \SI{5}{\kilo\electronvolt}, with a damage depth of \SI{1.9 \pm 1.2}{\nano\metre} for the sidewall and \SI{3.9 \pm 2.0}{\nano\metre} for the top surface. These simulations predict an implantation layer significantly smaller than the \SI{40}{\nano\metre} magnetically inactive layer estimated by Longenecker et al.\ using nuclear-spin MRFM experiments \cite{Longenecker2012nov}. It is worth noting that the SRIM simulation does not account for effects such as redeposition, milling geometry, and thermal effects. Further measurements are needed to quantify the effect of ion implantation, vacancy formation, and redeposition on the magnetization of different magnetic materials.

The pre-milling protocol greatly improves upon the conventional FIB fabrication methods.\cite{Stipe2001dec,Mamin2005jul,Budakian2004jan,Overweg2015aug}  The transfer procedure also minimizes overall ion beam exposure of both magnet and cantilever by using electron-beam-induced deposition (EBID) for platinum adhesion. This adhesion procedure reduces contamination and gallium implantation compared to focused-ion-beam-induced deposition (FIBID) and to the previous chip-based fabrication method.\cite{Longenecker2011may} The electrostatic transfer mechanism itself avoids any need for ion-beam or electron-beam welding during pickup and manipulation, further reducing magnet exposure to beam damage.

Additional strategies are available to further reduce milling damage and redeposition effects.
These include optimizing milling parameters,\cite{Yoon2017sep, Kim2012feb} gas-assisted milling with \ce{XeF2},\cite{Ganczarczyk2011jan} hybrid approaches using gallium for bulk material removal followed by low-energy argon or neon polishing,\cite{Kato2004oct, Xia2021feb} and physical protection with capping layers or masking materials.\cite{Thompson2006jan,Schaffer2012mar, Vermeij2018mar}

The transfer protocol is also constrained by the particle size. 
We found that the transfer yield was higher if the magnet particle and the cantilever-probe tip interaction area were of the same order of magnitude, around \SI{1}{\square\micro\metre}. A smaller particle could nevertheless be attached using a tip with a larger taper or by pre-milling the probe and cantilever tip to a smaller size;\cite{Campanella2011dec} larger particles, beyond those demonstrated here, are similarly accessible by scaling the probe tip size. To promote transfer in future experiments, the probe tip could be electrostatically biased to adjust the electrostatic force during both pickup and transfer to the cantilever.\cite{Giannuzzi2023feb, Stricklin2023jul}

Although developed for MRFM magnetic tips, our electrostatic transfer protocol reduces ion-beam and electron-beam damage and allows precise control over tip geometries and materials. The electrostatic transfer and attachment procedures introduced here are also versatile, enabling tailored operation for particles with varying materials, sizes, and shapes.
 
\section{\label{sec:Conclusion}Conclusion}

We developed a focused-ion-beam-assisted electrostatic transfer method to attach magnetic particles onto microcantilevers for magnetic resonance force microscopy. Demonstrated with \SI{460}{\nano\metre} diameter Ni spheres and \SI{2.8}{\micro\metre} long NdFeB cylinders, the procedure uses an FIB system with a tungsten probe to electrostatically transfer prefabricated particles onto a cantilever, securing them with electron-beam-induced platinum deposition.

The method addresses key limitations of previous fabrication methods by having compatibility with magnetic powders of various compositions, protecting the magnet's leading edge from direct ion irradiation, achieving precise, geometry-controlled magnet overhang to suppress electrostatic and noncontact friction noise, and providing geometric freedom for high-aspect-ratio structures. The geometric freedom and electrostatic transfer technique allow FIB milling procedures that minimize ion-beam damage. The method will enable a systematic investigation of how magnet shape, size, and material affect MRFM sensitivity, which is critical for understanding signal reduction in small-magnet electron spin resonance experiments.\cite{Boucher2023jan}  The technique introduced here is broadly applicable to other scanning probe methods needing custom-tipped probes, such as atomic force microscopy \cite{Wang2014aug, Ong2007jun}, and tip-enhanced Raman spectroscopy \cite{Stockle2000feb,Pettinger2004mar}. 
 
\section{\label{sec:Acknowledgements}Acknowledgments}

This work was supported by Cornell University and by the National Institute of General
Medical Sciences of the National Institutes of Health under
Award Number R01GM143556. This work made use of the Cornell Center for Materials Research shared instrumentation facility and Helios FIB acquisition, supported by the NSF (DMR-2039380).
This work was performed in part at the Cornell NanoScale Facility, a member of the National Nanotechnology Coordinated Infrastructure (NNCI), which is supported by the National Science Foundation (Grant NNCI-2025233).
We thank Michael C.\ Boucher for assistance with SRIM simulations. 
\bibliography{bib/references.bib}\end{document}